# Emergence of Rashba splitting and spin-valley properties in Janus MoGeSiP$_2$As$_2$ and WGeSiP$_2$As$_2$ monolayers

Ghulam Hussain[a,*], Abdus Samad[b], Majeed Ur Rehman[c,*], Giuseppe Cuono[a], Carmine Autieri[a]

[a]International Research Centre MagTop, Institute of Physics, Polish Academy of Sciences, Aleja Lotnikow 32/46, PL-02668 Warsaw, Poland

[b]Department of Physics and Energy Harvest Storage Center, University of Ulsan, Ulsan 44610, South Korea

[c]College of Physics and Optoelectronic Engineering, Shenzhen University, Shenzhen, Guangdong 518060, China

E-mail: ghussain@ifpan.edu.pl, majeed@mail.ustc.edu.cn

## ABSTRACT

First-principles calculations are performed to study the structural stability and spintronics properties of Janus MoGeSiP$_2$As$_2$ and WGeSiP$_2$As$_2$ monolayers. The high cohesive energies and the stable phonon modes confirm that both these structures are experimentally accessible. In contrast to pristine MoSi$_2$P$_4$, the Janus monolayers demonstrate reduced direct bandgaps and large spin-split states at K/K'. In addition, their spin textures exposed that breaking the mirror symmetry brings Rashba-type spin splitting in the systems which can be increased by using higher atomic spin-orbit coupling. The large valley spin splitting together with the Rashba splitting in these Janus monolayer structures can make a remarkable contribution to semiconductor valleytronics and spintronics.

**Introduction**

Two-dimensional (2D) structures of transition metal chalcogenides have been the subject of extensive study and have found tremendous applications in electronics and optoelectronics.[1-9] 2D transition metal dichalcogenides TMDCs have electronic bandgaps in the range of 1.1-1.9 eV, and reveal promising electronic and mechanical features, making them applicable in electronic devices and photonics.[10-17] Besides, the lack of inversion symmetry and large spin-orbit coupling (SOC) generated from d-orbitals of transition metals in TMDCs yield enormous spin splitting at the corners of hexagonal Brillouin zone.[13, 18, 19] The TMDCs are considered ideal materials in valleytronics due to the strong coupling amongst spin and valley degrees of freedom.[10, 20, 21] Also, the electronic and optical properties could be modulated using the alloy composition in TMDCs, for example in $Mo_xW_{1-x}S_2$, $MoS_xSe_{2-x}$, and $WS_{2x}Se_{2-2x}$.[22-26] In recent times, the Janus structure of MoSSe has been successfully grown in the 2H phase via chemical vapor deposition (CVD) method,[27, 28] thereby replacing the Se layer with S atoms in the case of $MoS_2$, while the S layer with Se atoms in $MoSe_2$. Different from $MX_2$ (M=Mo, W and X=S, Se) which possesses mirror symmetry, the Janus phase $MXY$ (*M*=Mo, W, *X*=S, Se, and *Y*=Se, S) shows an out-of-plane electric field owing to the broken mirror symmetry, consequently, the structure exhibits out-of-plane piezoelectricity and Rashba spin splitting.[29-33]

Recently discovered 2D materials such as $MoSi_2N_4$ and $WSi_2N_4$,[34] with outstanding mechanical and semiconducting properties, have stimulated great research interest in studying their other counterparts, $MA_2Z_4$ (where *M* stands for transition metal such as Mo and W, *A* for Si or Ge and *Z* represents N, P, or As).[35-44] The studies revealed remarkable electronic, thermal, mechanical, magnetic, and optoelectronic properties.[38, 44-55] Furthermore, the breaking of inversion symmetry, together with strong SOC, induces spin-split states, for instance, large spin-splitting has been observed in the valence bands of $MoSi_2N_4$, $WSi_2N_4$, $MoSi_2As_4$, $CrSi_2P_4$, and $CrSi_2N_4$ at K/K' valleys of the two-dimensional hexagonal Brillouin zone.[56-59] Nonetheless, the Janus phase that is likely to reveal Rashba spin-splitting in this class of materials is essential for spintronic applications.

In the present work, the Janus $XGeSiP_2As_2$ ($X$=Mo, W) monolayers are investigated using the relativistic density functional theory. We calculate the phonon spectra and the orbital-projected electronic band structures to study the structural stability and electronic properties of these systems. Together with the band structures, the in-plane and out-of-plane spin textures are obtained to address the spin-split states observed at K/K', and the Rashba-type spin splitting at the Γ-point in Janus monolayers. Contrary to pristine $MoSi_2P_4$, the Janus $XGeSiP_2As_2$ ($X$=Mo, W) monolayers demonstrate reduced direct bandgaps, large spin-split states and Rashba splitting, due to the broken mirror symmetry. We found enhanced Rashba splitting as compared to Janus monolayers of transition metal dichalcogenides, *MXY* (where *M* represents Mo or W and *X*≠*Y* denote S, Se, or Te),[30] which could be very promising in spintronics devices, particularly in Datta-Das spin field effect transistors.[60]

**Computational details**

First-principles relativistic calculations are carried out on the basis of density functional theory (DFT) using Vienna *Ab Initio* Simulation Package (VASP).[61, 62] The Perdew–Burke–Ernzerhof in the framework of generalized gradient approximation is used to treat electron exchange-correlation.[63] The projector augmented wave method is adopted to resolve the DFT Kohn-Sham equations via the plane-wave basis set. An energy cutoff of 350 eV is considered for plane-wave expansion of wave functions and for *k*-point sampling the Monkhorst–Pack scheme is utilized. A dense mesh of 15×15×1 *k*-point is used. Slab models are built using a vacuum layer of 20 Å in the out-of-plane direction. A force convergence criteria of 0.0001 eV/Å and energy tolerance of $10^{-7}$ eV are set for the lattice relaxation. 4×4×1 supercells of the Janus MoGeSiP$_2$As$_2$ and WGeSiP$_2$As$_2$ monolayers are considered for the phonon dispersion spectra calculations using the PHONOPY code.[64]

**Results and discussion**

The pristine MoSi$_2$P$_4$ monolayer is observed to crystallize in a hexagonal structure with space group P$\bar{6}$m2 (No. 187). Monolayer MoSi$_2$P$_4$ is seven-atom thick, strongly bonded, with atoms stacking order as P-Si-P-Mo-P-Si-P which can be seen as a sandwich of MoP$_2$ layer between the two Si-P layers (Fig. 1a). This structure preserves the mirror-plane symmetry with respect to Mo atom but breaks the spatial inversion symmetry. Figure 1b illustrates the projected band structure for pristine monolayer MoSi$_2$P$_4$ where a direct bandgap of 0.61 eV appears at the K-point between the valence band maximum (VBM) and conduction band minimum (CBM). Due to the broken spatial inversion symmetry, a spin splitting of approximately 137 meV can be seen in the VBM ($\lambda_{Kv}$), while in CBM ($\lambda_{Kc}$) it is ~3.9 meV. Contrary to K-point, the bands at the M- and Γ-points are two-fold spin degenerate, see supplementary Fig. S2. From Fig. 1b, it is evident that CBM at K-point is mainly contributed by Mo-$d_{z^2}$ orbitals, whereas the VBM is composed of Mo-$d_{xy}$ and Mo-$d_{x^2-y^2}$ states. In Fig. 1c, the side and top views of Janus MoGeSiP$_2$As$_2$ monolayer are presented, this shows the breaking of mirror symmetry with respect to the Mo plane and the stacking order as As-Si-As-Mo-P-Ge-P. As indicated, the broken symmetry results in unequal interatomic distances, for instance, the bond lengths of Mo-P, Mo-As, Si-As and Ge-P are 2.45, 2.54, 2.32 and 2.30 Å, respectively. Similar variations in the interatomic distances are also observed for WGeSiP$_2$As$_2$. The optimized lattice constant for MoGeSiP$_2$As$_2$ monolayer is calculated to be 3.525 Å, while that of WGeSiP$_2$As$_2$ is 3.564 Å.

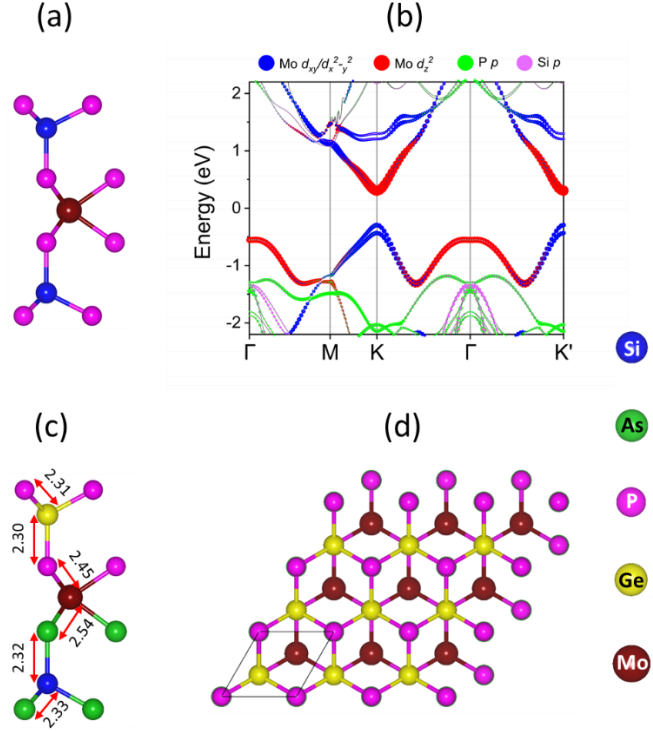

**Figure 1** (a) Side view of MoSi$_2$P$_4$ monolayer. (b) The projected bandstructure of MoSi$_2$P$_4$, the different colors represent the contribution of various orbitals to the wave functions at high symmetry k-points. The size of the dot is related to the amount of contribution, the bigger the size of the dot, the more the orbitals contribute. (c, d) Side and top views of Janus MoGeSiP$_2$As$_2$ monolayer with bond lengths shown in Å.

To study the stability of these Janus structures, we calculated the cohesive energies and performed the phonon calculations using 4×4×1 supercell. The cohesive energies ($E_c$) are calculated from ref. 58 and 59; such as for MoGeSiP$_2$As$_2$, the $E_c = E_{MoGeSiP_2As_2} − (E_{Mo} + E_{Ge} + E_{Si} + E_P + E_{As})$, where $E_{MoGeSiP_2As_2}$, and the energy terms in parenthesis ($E_{Mo}$, $E_{Ge}$, $E_{Si}$, $E_P$ and $E_{As}$) designate the total energy of monolayer MoGeSiP$_2$As$_2$ and that of individual atoms. They are calculated to be -2.77 eV/atom for MoGeSiP$_2$As$_2$, and -2.84 eV/atom in the case of WGeSiP$_2$As$_2$, which are higher than the previously reported Janus structures of MoSSe (−2.34 eV) and WSSe (-2.06 eV)[31]. These values are large enough to promise the stability of Janus structures in this class of 2D materials. Moreover, the phonon spectra are presented in Fig. 2, which is useful to investigate the dynamical stability of materials. We calculated the phonon band structures for $MoGeSiP_2As_2$ and $WGeSiP_2As_2$ along the Brillouin zone's high symmetry directions (K-Γ-M-K) using the finite difference method that is implemented in Phonopy code (Fig. 2a and 2b). The phonon spectra of both the $MoGeSiP_2As_2$ and $WGeSiP_2As_2$ reveal no negative frequency modes, indicating the dynamical stability of these Janus monolayers. On the other hand, the phonon band structures of $MoSnSiP_2As_2$ and $MoPbSiP_2As_2$ (shown in supplementary Fig. S1) exhibit the negative frequency phonon modes along the high-symmetry directions in their first Brillouin zone, hence they are dynamically unstable. From this, we infer that the Janus structures of $MoGeSiP_2As_2$ and $WGeSiP_2As_2$ can be experimentally accessible.

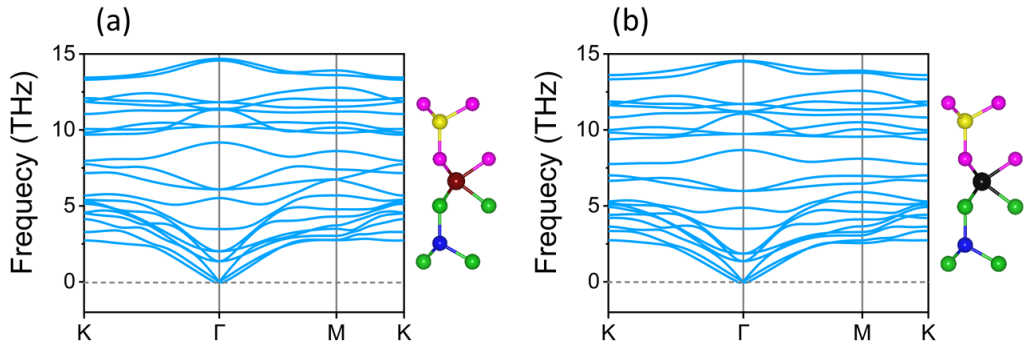

**Figure 2** Phonon spectra of Janus (a) $MoGeSiP_2As_2$, and (b) $WGeSiP_2As_2$ monolayers. Both the structures demonstrate positive phonon modes with no imaginary frequencies, confirming their dynamical stability.

We illustrate the planar average of the electrostatic potential energy in Fig. 3a where the energy difference (ΔΦ = 0.4 eV) represents the work function difference [65]. For the monolayer MoGeSiP$_2$As$_2$, the broken mirror symmetry with respect to the Mo-plane gives rise to a potential gradient normal to the basal plane. To better understand the electronic properties of MoGeSiP$_2$As$_2$, we calculated the orbital-projected band structures at PBE level considering the SOC as presented in Fig 3b. Both the VBM and CBM appear at the K-point of the Brillouin zone, indicating that the Janus phase retains the direct bandgap, however, with a slight decrease in the bandgap value (0.57 eV) with respect to pristine MoSi$_2$P$_4$ monolayer. Fig. 3b depicts that the CBM is principally composed of Mo-$d_z^2$ orbitals whereas the VBM is composed of Mo-$d_{xy}$ and Mo-$d_{x^2-y^2}$ states. Besides, the Zeeman-type spin splitting in the valence band, $\lambda_{Kv}$ is enhanced to 161.7 meV, while that of $\lambda_{Kc}$ to 10.6 meV as compared to the pristine MoSi$_2$P$_4$. Figures 3c,d show the spin textures for the two spin-splitted valence bands due to SOC, computed for the entire Brillouin zone centered at Γ-point. The arrows illustrate the in-plane components of spin polarization, while the colors display the out-of-plane component of spin polarization. Because of the time-reversal symmetry, these systems are overall nonmagnetic, since the opposite out-of-plane components of spin polarization can be clearly seen at the time-reversal high symmetry K and −K points. We found clockwise rotation for the in-plane spin components of the upper valence band (Fig. 3c), while counterclockwise for the lower valence band (Fig. 3d). Due to time-reversal symmetry, the spin splitting would likely be opposite at the two time-reversed high symmetry points of the Brillouin zone, and thus demonstrate opposite in-plane spin patterns at K/K' point. Interestingly, while the spin texture shows opposite polarizations for the two splitted bands, the same chirality is observed at K/K' points for the in-plane spin components in a single valence band. Most importantly, the two-fold spin degeneracy is lifted at the Γ-point for the Janus MoGeSiP$_2$As$_2$ monolayer, where the valence bands split not only in energy but also in momentum space, giving rise to Rashba spin splitting (see the inset of Fig. 3b and supplementary Fig. S2). At Γ-point, the valence bands are mainly contributed by Mo-$d_z^2$ orbitals and are slightly composed of P-$p$ and Ge-$p$ orbitals, defining the Rashba spin splitting in MoGeSiP$_2$As$_2$. The symmetry of MoGeSiP$_2$As$_2$ becomes trigonal C$_{3v}$, which is different from that of MoSi$_2$P$_4$ with space group D$_{3h}$ due to the out-of-plane asymmetry I (I ≠ Z → −Z). In addition, the Rashba energy $E_R$, the momentum offset $K_R$, and the Rashba parameter $α_R$ are obtained at PBE+SOC level. $E_R$ is the energy change of split states, $K_R$ characterizes the shift of bands in momentum space at Γ-point (as indicated in the right-most panel of Fig. S2), and $α_R$ can be expressed as, $α_R$ = 2$E_R$/$K_R$, which is calculated to be ~0.50 eV Å for MoGeSiP$_2$As$_2$. All the calculated parameters such as electronic band gap $E_g$, $λ_{Kv}$, $λ_{Kc}$, and $α_R$ are summarized in Table 1. Around the Γ point, we observe a sequence of 6 regions where $S_z$ becomes alternatively positive and negative. This is a signature of a strong cubic spin splitting,[66, 67] which is complementary to purely cubic Rashba and Dresselhaus types. This effect is particularly strong in Fig. 3c and it becomes weaker for the WSiP$_2$As$_2$. Figure 3e illustrates the constant energy spin-resolved 2D contours for the spin texture centered at the Γ-point; the $S_x$, $S_y$ and $S_z$ components are calculated in the $k_x$−$k_y$ plane. The Rashba-type splitting of spin-up and spin-down states is distinctly manifested by red and blue curves, respectively. The 2D Rashba spin splitting for the two valence bands (spin-up and spin-down) produces spin-textures with clockwise and anti-clockwise spin directions. This trend can be observed almost perfectly for the in-plane $S_x$ and $S_y$ spin components with inner and outer branches of Rashba-type split valence bands. However, the feature is weakly accompanied by the out-of-plane $S_z$ component, as observed in WSSe[30] and BiTeCl.[68]

**Table 1.** Electronic bandgap $E_g$, the valence band splitting $\lambda_{Kv}$ at K-point, conduction band splitting $\lambda_{Kc}$ at K-point, and the Rashba parameter $\alpha_R$ (eV Å).

| Material | $E_g$ (eV) | $\lambda_{Kv}$ (meV) | $\lambda_{Kc}$ (meV) | $\alpha_R$ (eV Å) |
| --- | --- | --- | --- | --- |
| MoSi$_2$P$_4$ | 0.61 | 137 | 3.90 | 0.00 |
| MoGeSiP$_2$As$_2$ | 0.57 | 162 | 10.6 | 0.50 |
| WGeSiP$_2$As$_2$ | 0.25 | 472 | 20.6 | 0.52 |

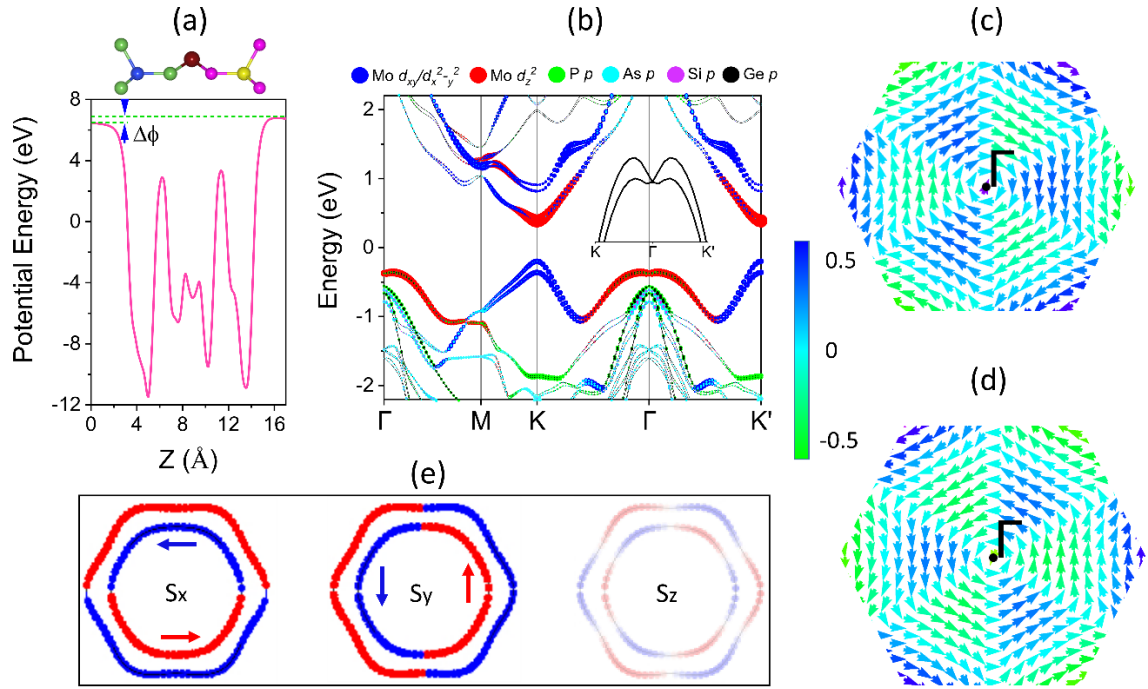

**Figure 3** (a) The planar average electrostatic potential energy of MoGeSiP$_2$As$_2$ monolayer, the work function difference, $\Delta\Phi$ is estimated to be ~0.4 eV. (b) Orbital-projected bandstructure of MoGeSiP$_2$As$_2$, the contribution of orbitals to the wave functions at high symmetry k-points is proportional to size of the colored dot i.e. the larger the dot size, the more they contribute. Also, the inset shows the zoomed illustration of the valence bands near the Γ-point to indicate Rashba spin splitting. (c, d) Spin textures of in-plane spin components for the upper and lower spin-splitted valence bands calculated for the complete BZ. The arrows represent the in-plane components of spin polarization, while the colors designate the out-of-plane component of spin polarization. (e) Spin-resolved constant energy 2D contours exhibiting the Rashba-type splitting for the spin projections i.e. S$_x$, S$_y$, and S$_z$, respectively.

To explore further the Janus structures in this family of materials, we also consider WGeSiP$_2$As$_2$, which essentially bears the same physics as MoGeSiP$_2$As$_2$, and attains the similar trigonal C$_{3v}$ symmetry in Janus form. Figure 4a displays the planar average of the electrostatic potential energy, which shows the work function difference, ΔΦ of ~0.45 eV. The orbital-projected bandstructure for WGeSiP$_2$As$_2$ monolayer is demonstrated in Fig. 4b; the electronic bandgap is significantly reduced (0.25 eV) with respect to pristine MoSi$_2$P$_4$ and Janus MoGeSiP$_2$As$_2$, while preserving the direct nature of its bandgap. In addition, the spin splitting $\lambda_{Kv}$ and $\lambda_{Kc}$ are enhanced in WGeSiP$_2$As$_2$ to 472 and 20.6 meV, respectively. The spin textures computed for the complete Brillouin zone centering at Γ-point for the two spin-splitted valence bands are shown in Fig. 4c,d. A clockwise (anticlockwise) rotations for the in-plane spin components of the upper (lower) valence bands are manifested. On the other hand, the out-of-plane spin components are represented by colors. Also, the spin splitting is opposite at the two time-reversed high symmetry K/K' points attributed to time-reversal symmetry. Similar to MoGeSiP$_2$As$_2$, while the two splitted bands show opposite polarizations in the spin textures, the same chirality is shown at K/K' points in a single valence band for WGeSiP$_2$As$_2$. In addition, the Rashba spin splitting is larger in WGeSiP$_2$As$_2$ as compared to MoGeSiP$_2$As$_2$ (see supplementary Fig. S2 and Table 1). From the projected bandstructure as shown, the CBM is maximally composed of W-$d_z^2$ orbitals, where the VBM is contributed by W-$d_{xy}$ and W-$d_{x^2-y^2}$ states. Similar to MoGeSiP$_2$As$_2$ the W-$d_z^2$ orbitals and relatively small contributions from P-$p$ and Ge-$p$ orbitals at Γ-point are responsible for the Rashba spin splitting. The Rashba parameter $\alpha_R$ for WGeSiP$_2$As$_2$ is observed to be slightly larger in magnitude than MoGeSiP$_2$As$_2$ as shown in Table 1. Further, we show the constant energy spin-resolved 2D contours spin texture in Fig. 4e, calculated in the k$_x$−k$_y$ plane. The spin-up (red arrow) and spin-down (blue arrow) states reveal Rashba-type splitting, displaying spin-textures with clockwise and counterclockwise spin directions. It is noticed that the Rashba-type splitting is strongly contributed by in-plane $S_x$ and $S_y$ spin components and weakly contributed by the out-of-plane $S_z$ component. We conclude that trigonal $C_{3v}$ symmetry brings a net dipole moment[69] in WGeSiP$_2$As$_2$ and MoGeSiP$_2$As$_2$ monolayers, which play a crucial role in the Zeeman-type bands splitting. In Figs. 3c,d and 4c,d, the arrows characterize the in-plane components of spin polarization and the colors show the out-of-plane component of spin polarization. Because of the time-reversal symmetry, these systems are overall nonmagnetic, since the opposite out-of-plane components of spin polarization can be clearly seen at time-reversed K and −K points.[69] Moreover, the broken mirror symmetry assures to assist the Rashba-type spin splitting in these Janus systems. From Table 1, it is concluded that WGeSiP$_2$As$_2$ shows relatively larger band splitting ($\lambda_{Kv}$, $\lambda_{Kc}$) and the Rashba parameters due to the heavy nature of the W atom.

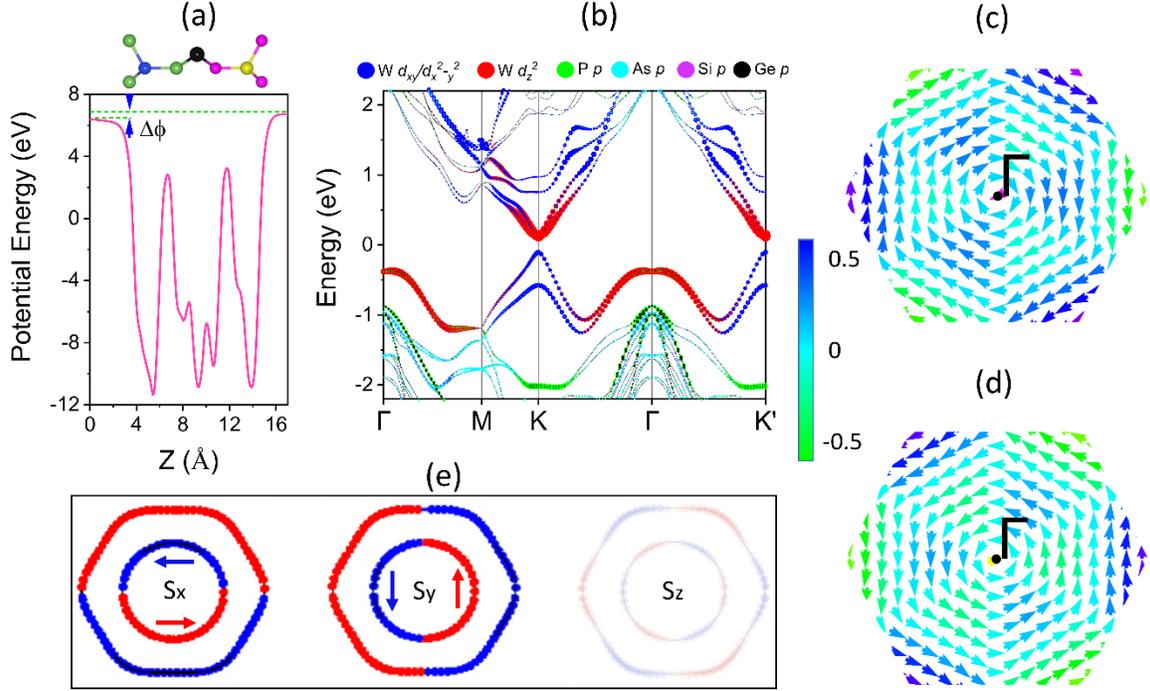

**Figure 4** (a) Planar average electrostatic potential energy of monolayer WGeSiP$_2$As$_2$, the work function difference, ΔΦ is approximated to be around 0.45 eV. (b) Orbital-projected bandstructure of WGeSiP$_2$As$_2$, the smaller the dot size, the less the orbitals contribute to wave functions and vice versa. (c, d) In-plane spin textures for the upper and lower spin-splitted valence bands centered at Γ-point. The arrows indicate the in-plane spin components of polarization, while the colors represent the out-of-plane component of spin polarization. (e) Constant energy 2D contours showing the Rashba-spin splitting for the S$_x$, S$_y$, and S$_z$ spin components.

**Conclusions**

In summary, we presented a detailed and systematic study on Janus $X$GeSiP$_2$As$_2$ ($X$=Mo, W) monolayers obtained via first-principles methods. The cohesive energies and phonon dispersion calculations demonstrated that these structures are stable. In electronic band structures, the intrinsic spin splitting at the K/K'-points of pristine MoSi$_2$P$_4$ is further enhanced in the Janus phase, suggesting their valleytronic applications. Moreover, the broken mirror symmetry in $X$GeSiP$_2$As$_2$ induces an out-of-plane electric field that results in Rashba-type spin splitting at the Γ-point. This splitting occurs due to the asymmetric potential in the Janus phase, perpendicular to the basal plane. Additionally, a signature of strong cubic spin splitting has been observed for the MoGeSiP$_2$As$_2$. Our study unveils the potential of Janus XGeSiP$_2$As$_2$ monolayers for valleytronics and spintronics.

**Supplementary material**

See supplementary material for spin splitting and phonon spectra of other Janus structures.

**Acknowledgements**

This work is supported by the Foundation for Polish Science through the international research agendas program co-financed by the European Union within the smart growth operational program. We acknowledge the access to the computing facilities of the Interdisciplinary Center of Modeling at the University of Warsaw, Grants No. G75-10, No. GB84-0, and No. GB84-7.